\documentclass[aps, prl, reprint, superscriptaddress, floatfix, longbibliography, nobibnotes]{revtex4-2}

\usepackage[utf8]{inputenc}
\usepackage{graphicx}
\usepackage{xcolor}
\usepackage{amsmath}
\usepackage{amsfonts}
\usepackage{amssymb}
\usepackage{siunitx}
\usepackage[colorlinks=true, citecolor=blue, urlcolor=red]{hyperref}
\usepackage{orcidlink}

\newcommand{\W}{\mathcal{W}}
\newcommand{\Q}{\mathcal{Q}}
\newcommand{\V}{\mathcal{V}}
\newcommand{\Teff}{T_{\textrm{eff}}}
\newcommand{\dd}{\mathrm{d}}
\newcommand{\erf}{\mathrm{erf}}

\begin{document}

\title{Fooling the Landauer bound with a demon biased thermal bath}
\author{Salambô Dago\,\orcidlink{0000-0002-1964-2375}}
\email[Corresponding author: ]{sdago@unistra.fr}
\altaffiliation{Current address: Institut de science et d'ingénierie supramoléculaires (ISIS) - UMR 7006, université de Strasbourg et CNRS}
\affiliation{CNRS, ENS de Lyon, Laboratoire de Physique, F-69342 Lyon, France}

\author{Ludovic Bellon\,\orcidlink{0000-0002-2499-8106}}
\affiliation{CNRS, ENS de Lyon, Laboratoire de Physique, F-69342 Lyon, France}

\date{\today}

\begin{abstract}
The Landauer principle establishes a fundamental lower bound on the energetic cost of the erasure for a one-bit memory in thermal equilibrium. Here, we experimentally demonstrate how this bound can be effectively circumvented by introducing an hysteresis in the feedback generated virtual potential of a micro-resonator acting as the information bit. By tuning the hysteresis, we engineer a non-equilibrium steady state with an adjustable effective temperature, enabling erasure processes that consume over $20\%$ below the Landauer bound. Our results reveal that the hysteresis acts as an embedded Maxwell demon, exploiting temporal and spatial information to reduce the system’s entropy and thermodynamic transformation's cost. This approach provides a versatile platform for exploring the interplay between feedback, information, and energy in stochastic systems.
\end{abstract}

\maketitle

\section{Introduction}
The thermodynamic cost of information processing has been widely studied for its fundamental implications and potential technological applications~\cite{Rex, Parrondo_sagawa, CiliLutz_PT, Orlov_book, Gammaitoni_2016, Konopik-2023, Toyabe2010, Parrondo2014, Pekola2014}. This cost is bounded from below by Landauer’s principle~\cite{Landauer_1961}, which states that at least $L_0 =\ln 2\, k_B T_0 $ of work is required to erase one bit of information from a memory in contact with a thermostat at temperature $T_0$, with $k_B$ the Boltzmann constant. Although this amount of energy is extremely small ($\sim \SI{3e-21}{J}$ at room temperature $T_0=\SI{300}{K}$), it represents a universal lower limit, reached asymptotically during quasi-static erasures and independent of the physical implementation of the memory. The Landauer bound $L_0$ has been verified experimentally in various platforms, including optical tweezers~\cite{Berut2012, Berut2015}, electrical circuits~\cite{orl12}, feedback traps~\cite{Bech2014, Gavrilov_EPL_2016, Proesmans-2020}, levitated optomechanics~\cite{Ciampini-2025}, micro-cantilevers in feedback-controlled virtual potentials~\cite{Dago-2021, Dago-2022}, nanomagnets~\cite{Hong_nano_2016, mar16}, and quantum systems such as trapped ions~\cite{Yan_2018} or molecular nanomagnets~\cite{gau17}.

To challenge the generality of the Landauer principle and explore the possibility of erasing information with a lower cost~\cite{Oriols-2023}, some underlying hypotheses in the description of the memory need to be relaxed. Notably, the effect of non-equilibrium features has been investigated on several levels~\cite{Esposito-2011}. To decrease the overhead to $L_0$ of finite-time erasures for instance, it has been theoretically proposed~\cite{Proesmans-2020, Proesmans-2020-PRE} and experimentally demonstrated~\cite{Oikawa-2025} to use out-of-equilibrium final states via adequate protocols. The Landauer bound itself can be challenged by preparing an out-of-equilibrium initial state, as suggested in Ref.~\onlinecite{Konopik-2020} and demonstrated with levitated optomechanics~\cite{Ciampini-2025}. A different approach consists in using an out-of-equilibrium thermostat: a squeezed thermal bath could potentially decrease the bound to zero~\cite{Klaers-2020}. For open quantum systems, the information backflow from the system to the environment leads to a memory effect and a non-Markovian dynamics, potentially altering (reducing or increasing) the Landauer bound~\cite{Chattopadhyay-2025}.

In this work, we extend previous studies performed with a classical underdamped micro-resonator used as a one-bit memory in a feedback-controlled virtual double-well potential~\cite{Dago-2022, Dago-2021}. The underdamped regime offers new insights on a wide variety of fundamental questions on the connections between feedback and thermodynamics~\cite{Seifert_2012, PhysRevLett.93.120602, PhysRevE.84.061110, rosinberg_stochastic_2015, PRLBarros}, the effect on entropy and information in the presence of a continuous feedback~\cite{Rosinberg_2016, Munakata_2012,Munakata_2013,Ch_trite_2019} introducing the concept of mutual information, or the theory around the Sagawa demon~\cite{PhysRevLett.104.090602}. When the outcome of random process is stored in an external agent (demon), systems can seemingly violate the second law of thermodynamics~\cite{Jar,phkv-wrsd}. As in the Szilard's engine~\cite{Szilard-1929}, the thermodynamic paradox is solved when taking into account the restoring of the external agent to its initial state~\cite{Bennett-1982}.

Here, we investigate how introducing a hysteresis in the feedback potential modifies the thermodynamic cost of information erasure. We show that this additional control scheme effectively shifts the system’s temperature, leading to an apparent increase or decrease of the system's temperature. Remarkably, under specific conditions, we observe erasure processes operating with an average energetic cost more than $20\%$ below the conventional Landauer bound. The gain in the erasure cost can be considered as the work of a demon (an intelligent creature able to monitor the system). This study thus provides new insight into the intimate relationship between feedback memory, information flow, and the thermodynamic limits of computation in stochastic systems.

The letter is organized as follows: we first present the experimental setup with a feedback controlled double well potential. Using a tunable hysteresis when switching between wells, we show how the Landauer bound for a 1-bit erasure can be fooled. We then model the system's kinetic temperature in a static double well potential, and propose a complete theoretical model that supports the experimental results. After underlining the validity of the observations beyond quasi-static erasure, we conclude by interpreting the results using a Maxwell demon description of the feedback.

\section{Experimental set-up} \label{hyst_exp}

\begin{figure}[htb]
 \includegraphics[width=\columnwidth]{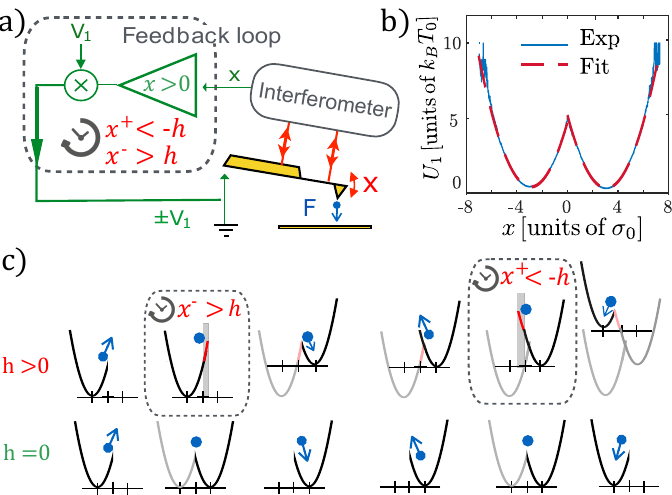}
\caption{\textbf{(a) Experimental setup.} The deflection $x$ of a conductive cantilever (yellow) is measured using a differential interferometer~\cite{Paolino2013}. The cantilever behaves as a harmonic oscillator whose equilibrium position is controlled electrostatically by the voltage difference with a facing electrode. The feedback controller (dashed box), composed of a comparator and a multiplier, generates a virtual double-well potential. In the standard configuration, the switch between voltages $\pm V_1$ occurs when $x=0$; with hysteresis, the switching threshold is shifted by $\pm h$~\cite{Dago-exp}.
\textbf{(b) Feedback-induced double-well potential for $h=0$.} Experimental potential (blue) reconstructed from the measured probability distribution function (PDF) of $x$ using the Boltzmann relation, in excellent agreement with the theoretical form $U(x)=\frac{1}{2}(|x|-x_1)^2$ (dashed red).
\textbf{(c) Energy pumping induced by the hysteresis.} For $h=0$ (bottom), the switching occurs exactly at $x=0$, so the system crosses smoothly between wells without accumulating potential energy. For $h>0$ (top), the delayed switch allows the cantilever to store additional potential energy, at the expense of losing some kinetic energy. This energy exchange occurs at every switching event biased by the ``demon'' hysteresis.
}
\label{schema_hyst}
\end{figure} 

The experiment uses a micro-cantilever of mass $m$, stiffness $k$, resonance angular frequency $\omega_0 = \sqrt{k/m} = 2\pi\times(\SI{1270}{Hz})$. The dynamics of the deflection $x$ of this mechanical resonator matches a simple harmonic oscillator with a quality factor $Q_f=10$ in air at atmospheric pressure and room temperature $T_0$, and a rest-position variance $\sigma_0^2 = k_BT_0/k \sim \SI{1}{nm^2}$. In the following, we use $\sigma_0$ as the unit length, $\omega_0^{-1}$ as the unit time, $k_BT_0$ as the unit energy, and $T_0$ as the unit temperature. Applying a voltage difference between the conducting cantilever and a facing electrode creates an electrostatic force displacing the equilibrium position to $x_1$ [Fig.~\ref{schema_hyst}(a)], with the corresponding quadratic single well potential $U_S(x,x_1)=\frac{1}{2}(x-x_1)^2$.

The underdamped memory is implemented by embedding this oscillator in a \textit{virtual double-well potential} generated by feedback control~\cite{Dago-exp}. This is achieved by dynamically changing the sign of $x_1$ each time the cantilever crosses the midpoint $x = 0$: the feedback instantly switches the potential center to the opposite side ($+x_1 \leftrightarrow -x_1$), effectively creating two distinct wells. The resulting effective double well potential is $U_D(x, x_1) = \frac{1}{2}(|x|-x_1)^2$, as demonstrated in Fig.~\ref{schema_hyst}(b). Using a large value $x_1=X_1=5$ imposes a large barrier $B=\frac{1}{2}X_1^2=12.5$ between the wells, ensuring long term stability of each metastable equilibrium. This experiment thus creates an underdamped memory, whose logical state (0 or 1) is mapped to the position $x$ of the cantilever in the potential (sign of $x$, left or right well).

Information processing is performed by applying a time protocol on the potential $U(x,t)$. A 1-bit erasure to reset the memory to state 0 in a time $2\tau$ is sketched in Fig.~\ref{Whysteresis}(a)~\cite{Dago-2021,Dago-2022}: starting from and returning to $U_D(x, X_1)$, we apply 
\begin{equation}
U(x,t) =
\begin{cases}
U_D\big[x,X_1\!\left(1 - t/\tau\right)\big], \quad t < \tau, \\[4pt]
U_S\big[x,X_1\!\left(1-t/\tau\right)\big], \quad \tau < t < 2\tau.
\end{cases}
\label{Uprotocol}
\end{equation}
To explore the quasi-static regime, we typically set $\tau=\SI{1}{s}\approx 8000\,\omega_0^{-1}$, much longer than the oscillator relaxation time $\tau_\mathrm{relax} = 2Q_f\omega_0^{-1}=\SI{2.5}{ms}$. The cost of information processing is measured through the stochastic work and heat $\W$ and $\Q$ defined in Ref.~\onlinecite{sek10} and used in previous publication~\cite{Dago-2021}. 

Using the feedback controller (implemented with a Fast Programming Gate Array), we introduce a feedback bias via a comparator with hysteresis. The potential switches from the lower to the upper well ($-x_1 \rightarrow +x_1$) when $x>h$, and conversely from the lower to the upper ($+x_1 \rightarrow -x_1$) when $x<-h$. Unlike a threshold at $x=0$ regardless of the crossing direction, hysteresis delays or anticipates the switching event. For $h>0$, the system needs to overshoot the barrier at each crossing, enhancing the stability of the 1-bit information. For $h<0$, switching is anticipated, which is enabled by enforcing a temporal lockout preventing successive switches within a quarter of the oscillation period~\cite{Dago-exp}. Such negative hysteresis is kept below $0.1$ to ensure reliable encoding. The hysteresis affects the system's thermodynamics only during information manipulation, as will be discussed in the following.

\section{Results: Beating the Landauer bound}

\begin{figure}[!htb]
 \includegraphics[width=\columnwidth]{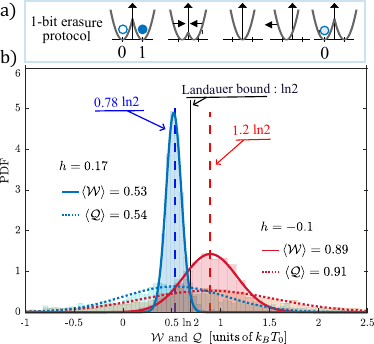}
 \caption{ \textbf{(a) Sketch of the erasure to $0$ protocol}. The two wells are first merged in a time $\tau$, the resulting single well is translated to the left in the same time $\tau$, and finally the initial double well is restored. Any initial condition end in logical state $0$.
 \textbf{(b) Work and heat distributions for quasi-static erasures performed with hysteresis parameters $h=0.17$ and $h=-0.10$}. The protocol is the one of Eq.~\ref{Uprotocol} with duration $\tau=\SI{1}{s}$. The average heat (PDF in dotted lines) perfectly matches the average work (PDF in plain lines) in both cases. For $h=0.17$ (blue), 1-bit erasures -- on the cooled down system -- require on average $78\%$ of the Landauer bound (dashed blue vertical line): $\langle \W \rangle =0.53 \pm 0.005$ and $\langle \Q \rangle= 0.54 \pm 0.02$. For $h=-0.10$ (red), the average energy cost (dashed red vertical line) -- on the warmed up system -- reaches $120\%$ of the Landauer bound: $\langle \W \rangle= 0.89\pm 0.015$ and $\langle \Q \rangle =0.91 \pm 0.02$.} 
 \label{Whysteresis}
\end{figure} 

Using this experimental setup with no hysteresis ($h=0$), quasi-static erasures require on average the Landauer bound for the work and for the heat~\cite{Dago-2021}: $ \langle \W \rangle = \langle \Q \rangle = L_0$. In this article we perform the same quasi-static erasures, introducing an hysteresis affecting the feedback-potential. We obtain the work and heat distributions plotted in Fig.~\ref{Whysteresis}(b) with the expected respectively gaussian and exponential tailed distributions~\cite{Dago-2021}. The Landauer bound is beaten by $22\%$ when a switching delay is imposed by the positive hysteresis $h=0.17$: $\langle \W \rangle \approx \langle \Q\rangle \approx 0.54 = 0.78\,L_0$. On the contrary when switch is anticipated by a negative hysteresis $h=-0.10$, the bound is exceeded by $30\%$: $\langle \W \rangle \approx \langle \Q\rangle \approx 0.9=1.30\,L_0$. Let us emphasize that our experimental platform is the only one in the field to offer an independent measurement of the work and heat distributions~\cite{Dago-2021}.

In the following, we explain how the Landauer bound is effectively modified. In short, a positive hysteresis corresponds to an effective cooling of the thermal bath, while a negative hysteresis leads to an effective heating.

\section{Model: demon thermal bath}
 
 \begin{figure}[!htb]
 \begin{center}
 \includegraphics[width=\columnwidth]{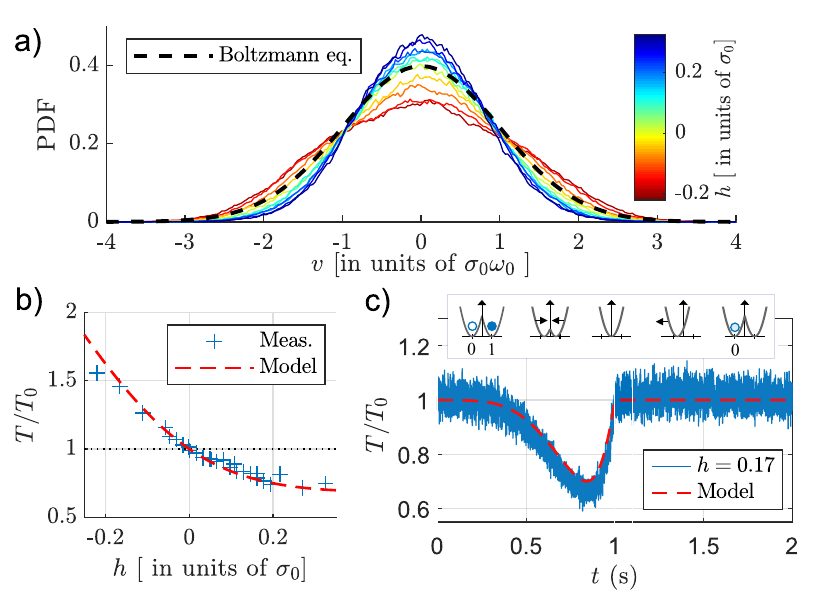}
 \end{center}
 \caption{\textbf{(a) Experimental PDFs of velocity for different hysteresis parameters $h$.} 
The velocity $\dot x$ is measured over $\SI{10}{s}$ of free evolution in a static double-well potential with centers in $\pm x_1=\pm1$ (chosen to maximize the effect), while varying the feedback hysteresis $h$. 
All PDFs remain Gaussian with variances (kinetic temperature) shown in (b) (blue crosses). For $h=0$, the Boltzmann equilibrium (normal distribution with variance $1$) is recovered (dashed black line).
\textbf{(b) Kinetic temperature $T=\langle \dot x^2 \rangle$ versus hysteresis parameter $h$.} Negative hysteresis increases the effective temperature (heating), whereas positive hysteresis decreases it (cooling), in excellent agreement with the theoretical prediction from Eq.~\ref{eqhyst} (dashed red). 
\textbf{(c) Temperature evolution during quasi-static erasure with feedback hysteresis $h=0.17$.} 
The temperature, extracted from the time-dependent velocity variance averaged over $N=2000$ trajectories, decreases during the first $\tau=\SI{1}{s}$ of the protocol due to the cooling effect of hysteresis in the virtual double-well. The amplitude of this cooling depends on the instantaneous well separation $x_1(t)$ and follows closely the theoretical model of Eq.~\eqref{eqhyst} (dashed red), as expected in the quasi-static regime.
}
 \label{rampehysteresis}
\end{figure} 

In this section, we show how the hysteresis $h$ creates a non-equilibrium steady state of effective temperature $T$, summarizing the results of Refs.~\onlinecite{Dago-exp,Dago-erratum}. This mechanism can be seen as closely related to the effective thermalization achieved through feedback cooling~\cite{gieseler_non-equilibrium_2015, gieseler_levitated_2018}.

Let us first notice that when switching between wells at a position $h \neq 0$, a work is performed on the system corresponding to the step in potential energy
\begin{equation}
\W = \Delta U_h = \frac{1}{2}\big[(h-x_1)^2 - (h+x_1)^2\big] = - 2 x_1 h.
\label{DeltaU}
\end{equation}
If $h>0$, energy is extracted from the oscillator, and on the contrary if $h <0$, energy is flowing toward the system. Such energy exchange occurs at each crossing of the barrier of height $B=\frac{1}{2}(x_1+h)^2$, we thus introduce the average barrier crossing rate $\Gamma(B/T)$ (see Ref.~\onlinecite{Dago-erratum} for the detailed expression, reminiscent of Kramers' escape rate for a system at temperature $T$ overcoming a barrier $B$) to deduce the mean potential energy flux $\langle \dot\W \rangle = \Gamma(\frac{1}{2}(x_1+h)^2/T) \Delta U_h$. This energy flux vanishes for small or large values of $x_1$ or $h$, since either $\Delta U_h$ or $\Gamma$ is small, and peaks for intermediate values: switches are frequent and energetic.

In a steady state, this constant work flux $\langle \dot\W \rangle$ performed by the demon triggering the potential switch needs to be equilibrated by the heat flux $\langle \dot\Q \rangle$ towards the thermostat. The latter can be related to the kinetic temperature $T=\langle \dot x^2\rangle$ of the system~\cite{Dago-2022}: 
\begin{equation}
\langle \dot\Q \rangle = \frac{1}{Q_f} (T-T_0) \label{eqbalancesimp}
\end{equation}
Combining the above with Eq.~\ref{DeltaU}, we obtain:
\begin{equation}
    \frac{1}{Q_f}(T-T_0)=-2 x_1 h \Gamma\left(\frac{(x_1+h)^2}{2T}\right) \label{eqhyst} 
\end{equation}
Eq.~\eqref{eqhyst} expresses the balance between the energy exchanged with the potential at each "delayed" switch, and the dissipation in the bath. The system temperature is the implicit solution of Eq.~\eqref{eqhyst} ensuring that these two fluxes compensate each other:
\begin{align} \label{solutioneqhyst}
 T=T_\mathrm{NESS}(x_1,h)  \text{ solution of Eq.~\eqref{eqhyst}}
\end{align} 
Eq.~\eqref{eqhyst} is solved numerically and compared with the experimental results in Fig.~\ref{rampehysteresis} which display the variances of the steady-state velocity distributions (kinetic temperature) for different values of the parameter $h$~\footnote{We focus here on the velocity PDF and how its variances is impacted by $h$, but the position PDF is similarly impacted and consists in the Boltzmann-like distribution at temperature $T$ instead of $T_0$. Nevertheless, as the PDF in the double well potential is complex, it is more immediate to look at the gaussian velocity PDF to study the consequences of the hysteresis.}. Model and experiment are in perfect agreement. Indeed, we experimentally observe that the velocity PDFs for $h<0$ and $h>0$ are respectively wider (higher variance, heats up) and sharper (lower variance, cools down) than the Boltzmann equilibrium ($h=0$).  We therefore demonstrate that system's effective temperature can be tuned between $T=0.7\,T_0$ and $T=1.55\,T_0$.

To model the temperature evolution during a quasi-static erasure, one simply needs to solve Eq.~\eqref{eqhyst} at each instant $t$, following the protocol $x_1(t)$ defined in Eq.~\eqref{Uprotocol}. It is important to note that this calculation is relevant only in the presence of a biased double-well potential, i.e., for $t < \tau$. For $t > \tau$, when the protocol imposes a single moving well, the system thermalizes with the bath, and the temperature returns to its equilibrium value: $T(t > \tau) = T_0$.

We compare in  Fig.~\ref{rampehysteresis}(c) the model $T(t)$ of Eq.~\ref{solutioneqhyst} (dashed line), to the experimental kinetic temperature (plain lines) - extracted from the velocity variance on thousands trajectories - during 1-bit erasures of parameters $\tau=\SI{1}{s}$,  $X_1=5$ and $h=0.17$. Experiment and model are again in very good agreement and show the expected cooling of the memory during the protocol.
The minimal temperature is reached for $x_1(0.85s)=0.75$ when the wells are nor too far apart (no crossing, $\Gamma \ll 1$) nor too close together (frequent switches but very little amount of energy pumped out, $\Delta U_h\ll 1$).

\section{Analysis: Landauer erasure in an effective thermal bath}

As demonstrated in Ref.~\onlinecite{Dago-2022}, when the memory temperature varies due to fast (non quasi-static) operations, the erasure cost is impacted accordingly and is worth $k_B\Teff\ln{2}$, where $\Teff$ is a weighted average of the kinetic temperature during the procedure. Although in Ref.~\onlinecite{Dago-2022} the temperature increase was induced by fast protocols, whereas here the temperature variation has its origin in $h\ne0$, the formalism remains the same and $\Teff$ writes,
\begin{align}
\Teff = \frac{1}{\ln 2} \int T\frac{\dd \ln{\V}}{\dd t}\left(1+\frac{\dd \ln T}{\dd \ln (x_1^2/T)} \right)\dd t, \label{Leff}  
\end{align}
with $\V=1+\erf(x_1/\sqrt{2})$, $T=T_\mathrm{NESS}(x_1,h)$, and $x_1(t)=X_1(1-t/\tau)$. For positive (negative) parameter $h$ we obviously expect $\Teff < T_0$ ($\Teff > T_0$).

Performing the numerical computation of Eq.~\ref{Leff} for the two hysteresis parameters of Fig.~\ref{Whysteresis} we extract the effective Landauer bounds $L_{\mathrm{eff}}=(\Teff/ T_0)  L_0$:
\begin{align}
L_{\mathrm{eff}}= 0.54 = 0.78\,L_0 &\text{ for } h=0.17 \label{LBh_pos} \\
L_{\mathrm{eff}}= 0.84 = 1.20\,L_0 &\text{ for } h=-0.10 \label{LBh_neg} 
\end{align}
This model prediction is again in excellent agreement with the experimental PDFs averages of Fig.~\ref{Whysteresis}.

\section{Finite time erasures}

\begin{figure}[!htb]
 \includegraphics[width=\columnwidth]{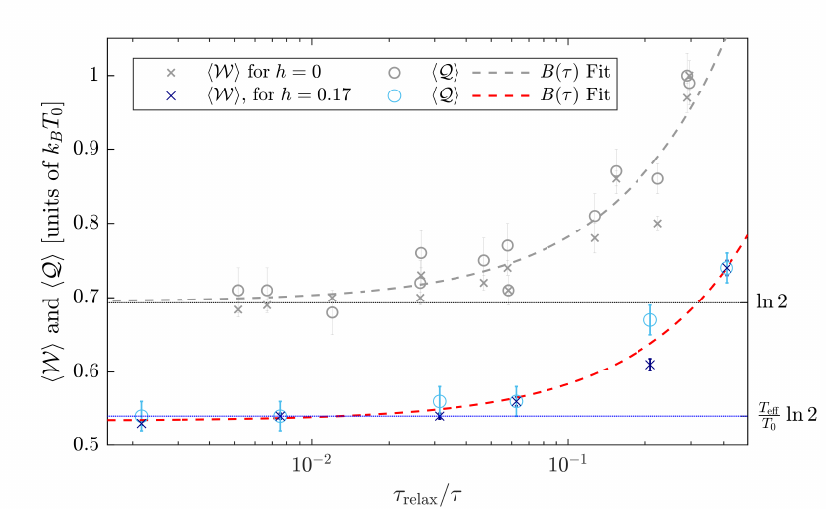}

 \caption{\textbf{Fast erasure cost} $\langle \W \rangle$ and $\langle \Q \rangle$ for erasure protocols plotted as a function of the usual $\tau_{\mathrm{relax}}/\tau$ scaling. Experimental data ($h=0.17$, blue markers) shows an asymptotic approach in $1/\tau$ (dashed red fit) to the effective Landauer bound $L_{\mathrm{eff}}= (\Teff/T_0) L_0$. The scaling of the energetic cost of fast erasures is similar to the case without hysteresis ($h=0$, grey markers and dashed line fit~\cite{Dago-2022}), only the asymptote differs.}
 \label{approachhysteresis}
\end{figure} 

So far, we have focused on quasi-static erasures to probe the minimal energetic cost under hysteresis feedback to challenge the Landauer bound. Faster erasures, with $\tau$ approaching $\tau_\mathrm{relax}$), are illustrated in Fig.~\ref{approachhysteresis} for $h = 0.17$. The energetics exhibit the usual $1/\tau$ scaling of the overhead to the Landauer bound (gray line, data from Ref.~\onlinecite{Dago-2022}), but this time converge towards the effective quasistatic  Landauer bound (colored curve). Fitting with the function $B(\tau)=\Teff/T_0 (a\, L_0+ b\, \tau_\mathrm{relax}/\tau)$, the best fit parameters are $a = 1.00 \pm 0.02$ and $b = 0.87 \pm 0.05$ for $h=0$ (with $\Teff=T_0$), and $a = 0.98 \pm 0.02$ and $b = 0.64 \pm 0.08$ for $h=0.17$ (with $\Teff=0.78\,T_0$). The similarity of these results further emphasize that operating with a biased virtual potential to implement the 1-bit memory effectively mimics a thermal bath of adjustable temperature for information processing.

\section{Conclusion}

We demonstrate in this work that introducing a small hysteresis into a feedback-controlled double-well memory enables its operation at an effective temperature, $T_{\mathrm{eff}}$. $T_{\mathrm{eff}}$ can be tuned below the bath temperature, allowing the system to effectively outperform the Landauer limit. We experimentally show that a 1-bit erasure can be achieved with an average work cost of $0.78\, k_{\mathrm{B}}T_0 \ln 2$. The results are in excellent agreement with our model for the steady-state temperature $T(x_1,h)$ in the virtual double-well, which captures the balance between the heat flux extracted by the feedback hysteresis and that dissipated into the thermal bath.

Nevertheless, the Landauer bound remains a universal constraint, intrinsically linked to the second law of thermodynamics. The apparent violation observed here originates from a Maxwell demon effectively embedded in the feedback loop, exploiting information stored during the threshold delay. In other words, the feedback “remembers” which well the system occupied previously and applies hysteresis accordingly when switching the potential. This stored information constitutes an entropy leak, explaining the reduced effective Landauer bound. Moreover, the inertia inherent to the underdamped dynamics is essential to enable this demon to reduce thermodynamic cost. We interpret by analogy with the Szilard engine~\cite{Neri-2025}: the temporal information stored in the system inertia  (e.g. "coming from the left") combined with spatial information of the system position used by the feedback comparator (crossing $\pm h$) reduce the entropy of the system in the double-well potential. As a consequence, the thermodynamic cost of this system used as a memory is also reduced (effectively lower temperature). This behavior also signs a non-Markovian dynamics, where the sign of $\pm x_1$ is the hidden variable needed to fully describe the system when $|x|<h$. In the limit $h \gg 1$ for example, the cantilever never escapes its initial well till $t=\tau$, where the information on the sign of $\pm x_1$ is erased by switching to a single well. The work measured in such a quasi-static protocol would be $0$, but the hidden cost of erasing the 1-bit information on the sign of $\pm x_1$ exactly $L_0$ !

Beyond this specific implementation, our results broaden the experimental and conceptual framework of information thermodynamics. They demonstrate how feedback control and memory can be harnessed to engineer non-equilibrium steady states with tunable effective temperatures, providing a versatile platform to study the energetic cost of information processing. Once properly understood, those results could be used to minimise computing thermodynamics in concrete devices, and enhance stochastic engines efficiency~\cite{PhysRevE.93.032146}. More generally, this approach opens new avenues in stochastic physics, where controlled feedback at the nanoscale may serve as a tool to probe the fundamental interplay between information, energy, and entropy.  

\acknowledgments
We thank Sergio Ciliberto for helpful discussions and collaboration on this project.

\subsection*{DATA AVAILABILITY} 
Datasets from the non-equilibrium steady-state analysis and standard Landauer erasures without deamon bias are already accessible respectively under Ref.\cite{dago_salambo_2022_6497247} and Ref.\cite{dago_salambo_2021_4807409}. The rest of the datasets used for this article will be uploaded soon on a dedicated Zenodo repository.

\bibliography{LandauerDemon}

\end{document}